\documentclass[aps,prl,twocolumn,showpacs,amsmath,amssymb,amsfonts,english]{revtex4}

\usepackage{babel}
\usepackage{graphicx}

\begin{document}


\title{Gravitational and electromagnetic fields near an anti-de~Sitter-like infinity}

\author{Pavel Krtou\v{s}}

\author{Ji\v{r}\'{\i} Podolsk\'y}

\affiliation{
  Institute of Theoretical Physics,
  Faculty of Mathematics and Physics, Charles University in Prague,\\
  V Hole\v{s}ovi\v{c}k\'{a}ch 2, 180 00 Prague 8, Czech Republic
  }

\date{October 17, 2003}  

\begin{abstract}
We analyze asymptotic structure of general gravitational and electromagnetic fields near
an anti-de~Sitter-like conformal infinity. Dependence of the
radiative component of the fields on a null direction along which the infinity is
approached is obtained. The directional pattern of
outgoing and ingoing radiation, which supplements
standard peeling property, is determined by the algebraic (Petrov) type
of the fields and also by orientation of principal null directions with
respect to the timelike infinity. The dependence on the orientation
is a new feature if compared to spacelike~infinity.
\end{abstract}


\maketitle

\longversion{\vspace*{-12pt}}


In spacetimes which are asymptotically flat the behavior of radiative
gravitational and electromagnetic
fields near infinity has been rigorously analyzed by means of now
classical techniques 
\cite{BondiBurgMetzner:1962,Penrose:1965,PenroseRindler:book2}.
However, it still remains an open
problem to fully characterize the asymptotic properties of
more general exact solutions of the
Einstein-Maxwell equations. Even in spacetimes
which admit a smooth infinity $\scri$ the concept of radiation is not
obvious when the cosmological constant $\Lambda$ is nonvanishing.
If we define \defterm{radiative component of field} as the $\afp^{-1}$ term
of the field with respect to a parallelly transported tetrad along
a null geodesic  ($\afp$ being affine parameter) then
for ${\Lambda\neq0}$  the radiation
depends on the direction along which geodesics approach
a given point at $\scri$ \cite{Penrose:1965,PenroseRindler:book2}.

It is natural to analyze and describe such dependence.
Recently, we studied \cite{KrtousPodolskyBicak:2003}
this behavior of fields near $\scri$ in
the case $\Lambda>0$ and demonstrated that  the directional pattern
of radiation close to de~Sitter-like infinity has a universal
character that is determined by the algebraic type of the fields. In the
present work we investigate the complementary situation when
$\Lambda<0$. Interestingly, although the method is similar to the previous
case, the results turn out to be more complicated, and completely new
phenomena occur. This stems from the fundamental difference that the
anti-de~Sitter-like infinity $\scri$ is \emph{timelike}, and thus admits a
\vague{richer structure} of radiative patterns. This fact was recently
demonstrated by analyzing radiation
generated by accelerating black holes in an anti-de~Sitter universe
\cite{PodolskyOrtaggioKrtous:2003}: $\scri$ is divided by the Killing
horizons into several domains with a different structure of
principal null directions, in which the patterns of radiation differ.
Moreover, ingoing and outgoing radiation have to be treated separately.
It is the purpose of our work to generalize these results and to
describe all the possible radiative patterns for gravitational and electromagnetic fields
near an anti-de~Sitter-like infinity.

A study of spacetimes with ${\Lambda\not=0}$
is motivated also by the fact that they have now become
commonly used in various branches of physical research, e.g. in
inflationary models, brane cosmologies, supergravity or string theories,
in particular due to the AdS/CFT correspondence.

\longversion{
\vspace{-12pt}
\section{Spacetime infinity, fields and tetrads}
\vspace{-9pt}
}

The \defterm{conformal infinity $\scri$} can be introduced
\cite{Penrose:1965,PenroseRindler:book2} as a boundary of
physical spacetime $\mfld$ with physical metric $\mtrc$,
when embedded into a larger conformal
manifold $\cmfld$ with conformal metric ${\cmtrc=\om^2\mtrc}$;
the \defterm{conformal factor $\om$} (negative in $\mfld$) vanishes on $\scri$.
Assuming $\cmtrc$ is regular there,
the metric $\mtrc$ is \vague{infinite} on $\scri$, and $\scri$ is thus
\emph{infinitely} distant from the interior of spacetime $\mfld$.
We will be interested here in \defterm{timelike} conformal infinity
which is characterized by a spacelike gradient $\grad\om$ on $\scri$.
The conformal metric $\cmtrc$~near such an anti-de~Sitter-like
infinity can always be decomposed into
Lorentzian 3-metric $\scrimtrc$ tangent to $\scri$, and a part orthogonal to it,
\begin{equation}\label{MtrcOnScri}
  \mtrc = \om^{-2} (\scrimtrc+\clapse^2\,\grad\om{}^2)\period
\end{equation}
Spacelike unit vector $\norm$ normal to the infinity is then
\begin{equation}\label{NormVect}
  \norm^{\mu}=-\om^{-1}\clapse\,\mtrc^{\mu\nu}\,\grad_\nu\om\period
\end{equation}

We denote the vectors of an \defterm{orthonormal tetrad}
as ${\tG,\,\qG,\,\rG,\,\sG}$ ($\tG$ timelike) and the associated null tetrad as
\begin{equation}\label{NormNullTetr}
\begin{aligned}
  \kG &= \textstyle{\frac1{\sqrt{2}}} (\tG+\qG)\comma&
  \lG &= \textstyle{\frac1{\sqrt{2}}} (\tG-\qG)\commae\\
  \mG &= \textstyle{\frac1{\sqrt{2}}} (\rG-i\,\sG)\comma&
  \bG &= \textstyle{\frac1{\sqrt{2}}} (\rG+i\,\sG)\commae
\end{aligned}
\end{equation}
so that $\kG\spr\lG\!=\!-1$, $\mG\spr\bG\!=\!1$.
In the null tetrad the Weyl tensor $\WT_{\alpha\beta\gamma\delta}$
can be parameterized by five complex coefficients $\WTP{}{j}$,\; ${j\!=\!0,\,1,\,2,\,3,\,4}$,
and the electromagnetic tensor $\EMF_{\alpha\beta}$ by three
coefficients $\EMP{}{j}$,\; ${j\!=\!0,\,1,\,2}$,
see \cite{Krameretal:book,KrtousPodolsky:2003a}.

We wish to investigate behavior of these field components in an
appropriate interpretation tetrad parallelly transported along
future oriented null geodesics $\geod(\afp)$ which
reach a given point $\scripoint$ at $\scri$.
Such geodesics form two distinct families which are distinguished
by their \defterm{orientation $\EPS$}:
geodesics \defterm{outgoing} to $\scri$
which \emph{end} at $\scripoint$ (${\EPS=+1}$),
and geodesics \defterm{ingoing} from $\scri$
which \emph{start} at $\scripoint$ (${\EPS=-1}$).
A geodesic thus reaches the point $\scripoint$
for the affine parameter ${\afp\to\EPS\,\infty}$. The
lapse-like function ${\clapse>0}$ and the conformal factor $\om<0$ \nopagebreak
can be expanded along the geodesic in powers of ${1/\afp}$ as \nopagebreak
${\clapse \lteq \clapse_\onscri + \dots}$,
${\om \lteq \EPS\, \om_* \afp^{-1} + \dots}$.
Here, ${\clapse_\onscri\!=\!\clapse|_{\scripoint}}$ is \nopagebreak
the same for all geodesics reaching $\scripoint$.
Moreover, we require that the approach of all geodesics
to the infinity is \vague{comparable}, independent
on their \emph{direction}, so we assume $\om_*$ to be a (negative) constant.
It is equivalent to fixing the momentum ${p_\refT=\tens{p}\spr\norm}$
(${\tens{p}=\frac{D\geod}{d\afp}}$ being 4-momentum)
at a given small value of $\om$.
This choice of the \vague{comparable} approach to $\scri$ is the only one
we can apply unless there are additional geometrical
structures (as, e.g., a Killing vector) which would allow us to
fix a different quantity (e.g., the energy).
We will see that this choice has significant consequences
for the character of the radiation pattern.

The \defterm{interpretation tetrad ${\kI,\,\lI,\mI,\,\bI}$} also has to be
specified \vague{comparably} for all geodesics having different directions.
We require that: (i) Null vector $\kI$ is proportional
to the tangent vector of the geodesic
\begin{equation}\label{kIdef}
  \kI=\frac1{\sqrt{2}\clapse_\onscri}\,\frac{D\geod}{d\afp}\commae
\end{equation}
the factor being independent of the direction.
(ii) Null vector $\lI$ is fixed by normalization ${\kI\spr\lI=-1}$ and
requirement that normal vector $\norm$ belongs to
$\kI\textdash\lI$ plane \cite{PenroseRindler:book2}. Remaining vectors ${\mI,\,\bI}$
cannot be specified canonically. Below, these vectors will
be chosen arbitrarily and we will only study moduli
$\abs{\WTP{\intT}{4}}$ and $\abs{\EMP{\intT}{2}}$
of the radiative field components
which are independent of such a  choice.

As ${\afp\to\EPS\,\infty}$, the interpretation tetrad is \vague{infinitely}
boosted with respect to an observer with 4-velocity tangent to $\scri$.
To see this explicitly, we introduce an auxiliary tetrad ${\tB,\,\qB,\,\rB,\,\sB}$
adapted to the infinity, ${\qB=\EPS\,\norm}$, with
timelike vector $\tB$ given by the projection of $\kI$ to~$\scri$,
\begin{equation}\label{BstTetrScri}
  \tB\propto
  \kI-(\kI\spr\norm)\,\norm\commae
\end{equation}
and the spatial vectors ${\rB,\,\sB}$ being identical to  ${\rI,\,\sI}$.
Checking that
${\kI\spr\norm\lteq{\textstyle\EPS\frac1{\sqrt{2}}}\,\afp^{-1}}$
we obtain
\begin{equation}\label{BstIntRel}
\begin{gathered}
  \kI = B_\intT\; \kB = \afp^{-1}{\textstyle\frac1{\sqrt2}}\,(\tB+\EPS\,\norm)\comma\mI=\mB\commae\\
  \lI = B_\intT^{-1}\, \lB = \;\;\afp\;{\textstyle\frac1{\sqrt2}}\,(\tB-\EPS\,\norm)\comma\bI=\bB\commae
\end{gathered}
\end{equation}
${B_\intT=1/\afp}$ being a boost parameter which approaches zero on $\scri$,
i.e., it represents an \vague{infinite} boost.
Under this the fields transform as
${\WTP{\intT}{j}=B_\intT^{2-j}\,\WTP{\bstT}{j}}$,
${\EMP{\intT}{j}=B_\intT^{1-j}\,\EMP{\bstT}{j}}$.
Considering the behavior \eqref{fieldsnearscri}
in a tetrad adapted to $\scri$ we obtain \emph{peeling-off} property.

\longversion{
\vspace{-12pt}
\section{Directional pattern of radiation}
\vspace{-9pt}
}

\begin{figure}
\includegraphics{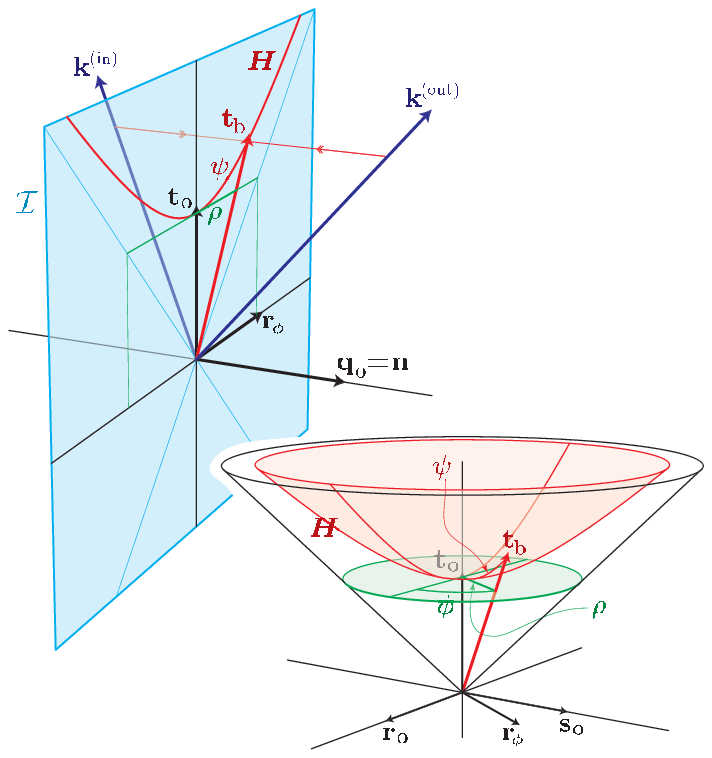}
\longversion{\vspace*{-3pt}}
\caption{\label{fig:nulldir}%
Parametrization of a null direction $\kG$
near timelike infinity $\scri$.
All null directions form three families:
\defterm{outgoing} directions (${\kG\spr\norm>0}$,
vector $\kG^{(\mathrm{out})}$ in the figure),
\defterm{ingoing} directions (${\kG\spr\norm<0}$,
vector~$\kG^{(\mathrm{in})}$), and directions tangent to~$\scri$.
With respect to
a reference tetrad ${\tO,\,\qO,\,\rO,\,\sO}$,
a direction $\kG$
can be parameterized by boost~$\PSI$, angle $\PHI$
and orientation $\EPS$,
or by parameters $\RHO$, $\PHI$,
or by a complex number~$\R$.
In the upper diagram, the vectors
${\tO,\,\qO,\,\rG_\PHI}$ are depicted, remaining
spatial direction $\sG_\PHI$ is suppressed; in the bottom
the direction ${\qO\!=\!\norm}$ is omitted.
The parameters ${\PSI,\,\PHI}$ specify the normalized orthogonal
projection $\tB$ of $\kG$ into $\scri$,
cf.\ Eqs.~\eqref{BstTetrScri}, \eqref{PSIPHIdef}.
To parametrize $\kG$ uniquely, we have to specify also
its orientation ${\EPS=\sign(\kG\!\spr\!\norm)}$ with respect to $\scri$.
Vectors~$\tB$ corresponding to all outgoing
(or ingoing) null directions form
a hyperbolic surface $\boldsymbol{H}$.
This can be radially mapped onto a two-dimensional disk tangent to
the hyperboloid at $\tO$,
which can be parametrized by angle $\PHI$ and radial coordinate ${\RHO=\tanh\PSI}$.
In the exceptional case ${\EPS=0}$ the boost ${\PSI\!\to\!\infty}$,
and ${\kG\propto\tB\!+\!\rG_\PHI}$ is tangent to $\scri$.
Finally, parameter $\R$ is the Lorentzian stereographic representation
of ${\PSI,\,\PHI,\,\EPS}$, cf.\ Eq.~\eqref{Rdef}.
}\longversion{\vspace*{-9pt}}
\end{figure}

Now we explicitly derive dependence of the radiation
on the direction of a null geodesic along which
the infinity is approached.
First, we parametrize this direction
with respect to a suitable
\defterm{reference tetrad}
${\tO,\,\qO,\,\rO,\,\sO}$ adapted to the conformal infinity, namely
${\qO=\norm}$.
The vectors $\tO,\,\rO,\,\sO$ can be fixed conveniently
with help of the particular geometry of the spacetime.
The timelike vector $\tB$
is related to the vector $\tO$ by a boost (cf.~Fig.~\ref{fig:nulldir})
\begin{equation}\label{PSIPHIdef}
  \tB = \cosh\PSI\;\tO + \sinh\PSI\;\rG_\PHI\commae
\end{equation}
with $\rG_\PHI\! =\! \cos\PHI\;\rO\!+\sin\PHI\;\sO$
(and $\sG_\PHI\! =\! -\!\sin\PHI\;\rO\!+\cos\PHI\;\sO$).
Because the vector $\tB$ is related to the projection of $\kI$
we can use the \vague{Lorentzian angles}
$\PSI$, $\PHI$ and the orientation $\EPS$
to parameterize the direction of the null geodesic.
Instead of these parameters it is also convenient to use their
\defterm{Lorentzian stereographic representation} $\R$,
\begin{equation}\label{Rdef}
  \R =
\begin{cases}
\tanh(\PSI/2)\,\exp(-i\PHI) &\;\text{for $\EPS=+1$}\commae\\[0.3ex]
\coth(\PSI/2)\,\exp(-i\PHI) &\;\text{for $\EPS=-1$}\period
\end{cases}
\end{equation}
We allow also the infinite value ${\R=\infty}$ corresponding to
${\PSI=0}$, ${\EPS=-1}$, i.e., ${\kG\propto\frac1{\sqrt2}(\tO-\qO)}$.

Next, we express the field components $\WTP{\refT}{j}$
(and $\EMP{\refT}{j}$) with respect to the reference tetrad
using algebraically privileged \defterm{principal null directions} (PNDs).
PNDs of gravitational (or electromagnetic, respectively) field
are null directions $\kG$ such that
${\WTP{}{0}=0}$ (or ${\EMP{}{0}=0}$)
in a null tetrad ${\kG,\,\lG,\,\mG,\,\bG}$
(a choice of ${\lG,\,\mG,\,\bG}$ being irrelevant).
If we parametrize $\kG$ by the above stereographic
parameter $\R$, 
the condition on PND with respect to the reference
tetrad takes the form \cite{Krameretal:book,KrtousPodolsky:2003a}
\begin{equation}\label{PNDcond}
\begin{gathered}
  \R^4 \WTP{\refT}{4} + 4 \R^3 \WTP{\refT}{3} +
    6 \R^2 \WTP{\refT}{2} + 4 \R\, \WTP{\refT}{1} + \WTP{\refT}{0}=0\commae\\
  \R^2 \EMP{\refT}{2} + 2 \R\, \EMP{\refT}{1} + \EMP{\refT}{0}=0\comma\text{respectively}\period
\end{gathered}
\end{equation}
There are thus four (or two) PNDs characterized by
the roots ${\R=\R_n}$, ${n=1,\,2,\,3,\,4}$ (or ${\R=\R^\EM_n}$, ${n=1,\,2}$).
In a generic situation we have ${\WTP{\refT}{4}\neq0}$,
and the remaining components
$\WTP{\refT}{j}$,\; ${j=0,\,1,\,2,\,3}$,
can be expressed in terms of $\R_n$
(analogously for $\EMP{\refT}{j}$,\; ${j=0,\,1}$),
see \cite{KrtousPodolskyBicak:2003}.

Using the conditions (i), (ii) above and
Eqs.~\eqref{BstIntRel}, \eqref{PSIPHIdef}, \eqref{Rdef},
we can now find the Lorentz transformation
from the reference tetrad to the interpretation
tetrad (up to a non-unique rotation
in the ${\mI\textdash\bI}$ plane).
We can thus
express the field components $\WTP{\intT}{4}$
(or $\EMP{\intT}{2}$) with respect to the interpretation tetrad
in terms of $\WTP{\refT}{j}$ (or $\EMP{\refT}{j}$),
and consequently in terms of the parameters $\R_n$ of PNDs
and $\WTP{\refT}{4}$
(or ${\R^\EM_n}$ and $\EMP{\refT}{2}$),
cf.\ \cite{KrtousPodolskyBicak:2003}.
Taking into account a typical behavior of the
fields in a tetrad adapted to $\scri$
(e.g., \cite{PenroseRindler:book2}),
\begin{equation}\label{fieldsnearscri}
  \WTP{\refT}{n}\lteq\WTP{\refT}{n}{}_*\;\afp^{-3}\comma
  \EMP{\refT}{n}\lteq\EMP{\refT}{n}{}_*\;\afp^{-2}\commae
\end{equation}
we finally obtain the \defterm{directional pattern
of radiation}---the dependence of radiative components of
gravitational and electromagnetic fields on the null direction
(given by~$\R$) along which the timelike infinity is approached:
\begin{align}
\begin{split}
\abs{\WTP{\intT}{4}} &\lteq \abs{\WTP{\refT}{4}{}_*}\,\afp^{-1}\,\bigabs{1-\abs{\R}^2}^{-2}\\
&\quad\quad\times{\textstyle
\abs{1-\frac{\R_1}{\R\mir}}\abs{1-\frac{\R_2}{\R\mir}}\abs{1-\frac{\R_3}{\R\mir}}\abs{1-\frac{\R_4}{\R\mir}}}\commae
\end{split}\label{WTPintT}\\
\begin{split}
\abs{\EMP{\intT}{2}} &\lteq \abs{\EMP{\refT}{2}{}_*}\afp^{-1}\bigabs{1-\abs{\R}^2}^{-1}
{\textstyle\abs{1-\frac{\R^\EM_1}{\R\mir}}\!\abs{1-\frac{\R^\EM_2}{\R\mir}}}\period
\end{split}\label{EMPintT}
\end{align}
Here, the complex number $\R\mir$,
\begin{equation}\label{mirdef}
  \R\mir = {\bar R}^{-1} = \coth^\EPS(\PSI/2)\,\exp(-i\PHI)\commae
\end{equation}
characterizes a direction obtained from the direction $\R$
by a \emph{reflection with respect to $\scri$}, i.e., the \defterm{mirrored}
direction with ${\PSI\mir=\PSI}$, ${\PHI\mir=\PHI}$ but opposite orientation ${\EPS\mir=-\EPS}$.

The expression \eqref{WTPintT} has been derived
assuming ${\WTP{\refT}{4}\neq0}$, i.e., ${\R_n\neq\infty}$.
However, to describe PND oriented along $\lO$
it is necessary to use a different component $\WTP{\refT}{j}$
as a normalization factor. E.g., with $\WTP{\refT}{0}$ we obtain
\begin{equation}
\begin{split}
\abs{\WTP{\intT}{4}} &\lteq \abs{\WTP{\refT}{0}{}_*}\,\afp^{-1}\,\bigabs{1-\abs{\R\mir}^2}^{-2}\\
&\quad\quad\times{\textstyle
\bigabs{1\!-\!\frac{\R_1{}\mir}{\R}}\bigabs{1\!-\!\frac{\R_2{}\mir}{\R}}\bigabs{1\!-\!\frac{\R_3{}\mir}{\R}}\bigabs{1\!-\!\frac{\R_4{}\mir}{\R}}}\period
\end{split}\label{WTPintTmir}
\end{equation}
Interestingly, the radiation pattern thus has the same form if we reflect
all PNDs, ${\R_n\to(\R_n)\mir}$, and switch ingoing and
outgoing directions, ${\R\to\R\mir}$.

\longversion{
\vspace{-12pt}
\section{Discussion}
\vspace{-9pt}
}

The expressions \eqref{WTPintT} and \eqref{EMPintT} characterize the asymptotic
behavior of the fields near anti-de~Sitter-like \nopagebreak
infinity. We will analyze here only gravitational field, \nopagebreak
discussion of electromagnetic one is analogous. First, we observe that
the radiation \vague{blows up} for directions with ${\abs{\R}\!=\!1}$
(i.e., ${\PSI\to\infty}$). These are null directions \emph{tangent} to
the infinity $\scri$, and thus they do not represent a direction
of any geodesic approaching the infinity from
the \vague{interior} of the spacetime. The reason for this divergent behavior
is purely kinematic: when we required the \vague{comparable} approach
of geodesics to the infinity
we had fixed the component of the 4-momentum $\mom\propto\kI$ normal to $\scri$.
Clearly, such a condition implies an \vague{infinite} rescaling if
$\kI$ is tangent to $\scri$ which results in the divergence of $\abs{\WTP{\intT}{4}}$.

\begin{figure}
\includegraphics{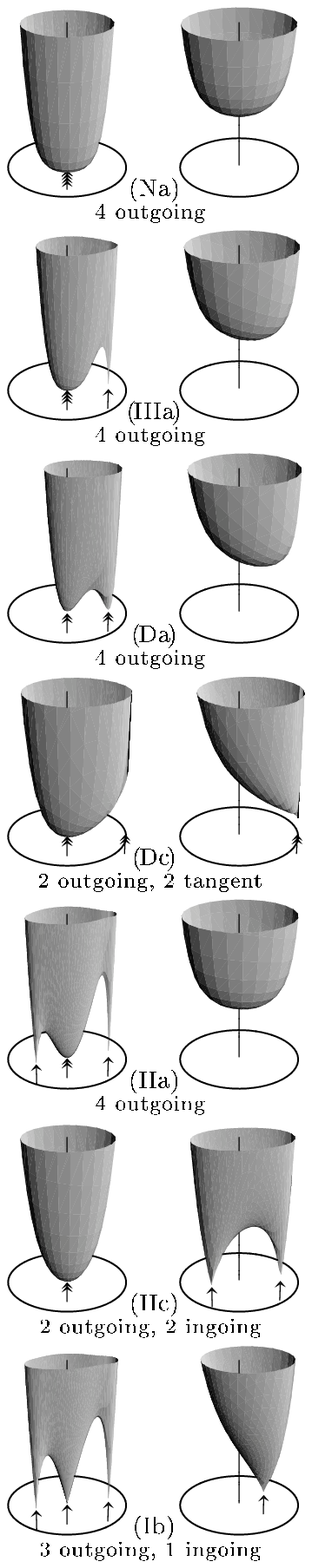}\;
\includegraphics{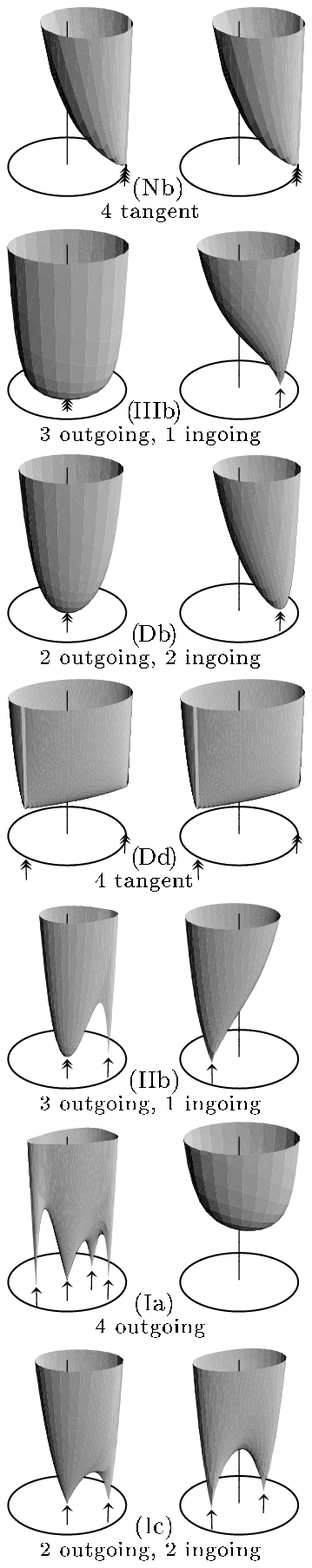}
\caption{\label{fig:dpr}%
Directional patterns of radiation near a timelike $\scri$.
All 11 qualitatively different shapes of the pattern
when PNDs are not tangent to $\scri$ are shown (remaining 9 are
related by a simple reflection with respect to $\scri$).
Patterns (Nb), (Dc), (Dd) are just few examples
with PNDs tangent to $\scri$.
Each diagram consists of patterns for ingoing
(left) and outgoing geodesics (right). $\abs{\WTP{\intT}{4}}$ is
drawn on the vertical axis, directions of geodesics
are represented on the horizontal disc by coordinates ${\RHO,\,\PHI}$
introduced in Fig.~\ref{fig:nulldir}.
\emph{Reflected}
[degenerated] PNDs are indicated by [multiple] arrows under the discs.
For PNDs that are not tangent to $\scri$ these are directions of vanishing radiation.
The Petrov type (N,~III, D, II, I) corresponding to the degeneracy of PNDs is
indicated by labels of diagrams, number of ingoing and outgoing PNDs is also displayed.
}
\end{figure}

The divergence at ${\abs{\R}\!=\!1}$ splits
the radiation pattern into two components---the pattern for
\emph{outgoing} geodesics (${\abs{\R}\!<\!1}$, ${\EPS=+1}$) and that
for \emph{ingoing} geodesics (${\abs{\R}\!>\!1}$, ${\EPS=-1}$).
These two different patterns are depicted in diagrams in Fig.~\ref{fig:dpr}
separately.

From Eq.~\eqref{WTPintT} it is obvious that
there are, in \mbox{general}, \emph{four} directions
along which the radiation \emph{vanishes}, name\-ly PNDs reflected with respect to
$\scri$, given by ${\R\!=\!(\R_n)\mir}$.
Outgoing PNDs give rise to zeros in
the radiation pattern for ingoing geodesics, and vice versa.
A qualitative shape of the
radiation pattern thus depends on
(i) \emph{orientation} of PNDs with respect to
$\scri$ (i.e., the number of outgoing/ingoing/tangent PNDs), and
(ii) \emph{degeneracy} of PNDs
(Petrov type of the spacetime).
Depending on these factors there are 51 qualitatively
different shapes of the radiation patterns
(3 for Petrov type N spacetimes, 9 for type III,
6 for D, 18 for II, and 15 for type I spacetimes);
21 pairs of them are related by the duality
of Eqs.~\eqref{WTPintT} and \eqref{WTPintTmir}.
The most typical are shown in Fig.~\ref{fig:dpr}.

The reference tetrad can be chosen to capture a geometry of the spacetime.
To simplify the radiation pattern we can
also adapt it to the algebraic structure, i.e.,
to correlate the tetrad with PNDs. For example, we can always orient $\tO$ along
the orthogonal projection to $\scri$ of the most degenerate
PND, say $\kG_4$. For outgoing $\kG_4$
we then obtain ${\kG_4\propto\kO}$, ${\R_4=0}$ (${\PSI_4=0}$, ${\EPS_4=+1}$);
for ingoing $\kG_4$ we get ${\kG_4\propto\lO}$, ${\R_4=\infty}$
(${\PSI_4=0}$, ${\EPS_4=-1}$)
and we have to employ the pattern \eqref{WTPintTmir}.
Thus, for spacetime of the Petrov type~N we get
$\PSI_n=0$, ${n=1,\,2,\,3,\,4}$, and the directional dependence
\begin{equation}\label{dprN}
\abs{\WTP{\intT}{4}}\propto(\cosh\PSI+\EPS_1\EPS)^2
\end{equation}
illustrated in Fig.~\ref{fig:dpr}(Na). Similarly, the radiation
pattern simplifies for other algebraically
special spacetimes.

At generic points the PNDs are not tangent to $\scri$.
However, they can be tangent on some lower-dimensional
subspace such as the intersection of $\scri$ with Killing
horizons---cf.\ anti-de~Sitter $C$-metric \cite{PodolskyOrtaggioKrtous:2003}.
These subspaces are important, e.g., in
the context of the Randall-Sundrum model:
a brane constructed from $C$-metric reaches the infinity
with PNDs tangent both to it and to $\scri$ \cite{Emparanetal:2000b}.

In the case when PND $\kG_1$ \emph{is tangent} to $\scri$, the reference
tetrad has to be chosen differently, e.g., in such a way that ${\R_4=1}$.
For \emph{type}~N spacetime we then obtain
the directional dependence
(see Fig.~\ref{fig:dpr}(Nb))
\begin{equation}\label{dprNtang}
\abs{\WTP{\intT}{4}}
  \propto\frac{\abs{1-\R}{}^{\!4}}{\bigabs{1-\abs{\R}{}^2}{}^2}
  =(\cosh\PSI-\sinh\PSI\cos\PHI)^2\period
\end{equation}
The only zero of this expression is for ${\R=1}$
(${\PSI\to\infty}$, ${\PHI=0}$; limit considered through directions with ${\abs{\R}\neq1}$)
which does not correspond to any outgoing or ingoing geodesic.
For \emph{type}~D spacetime (${\R_1=\R_2}$, ${\R_3=\R_4=1}$)
the directional dependence becomes
(Figs.~\ref{fig:dpr}(Dc), (Dd))
\begin{equation}\label{dprDtang}
\abs{\WTP{\intT}{4}}
  \propto\frac{\abs{1-\R}{}^{\!2}\abs{1-\R_1/R\mir}{}^{\!2}}{\bigabs{1-\abs{\R}{}^2}{}^2}\period
\end{equation}
This has zero at ${\R\!=\!(\R_1)\mir}$ (if ${\abs{\R_1}\!\neq\!1}$),
and it does \emph{not} diverge for
${\R=1}$, with a directionally dependent limit there.
If \emph{all PNDs are tangent} to $\scri$, ${\R_n=\exp(-i\PHI_n)}$, (not necessary degenerated) the
pattern can be written
\begin{equation}\label{dprtang}
\begin{split}
\abs{\WTP{\intT}{4}}\lteq \abs{\WTP{\refT}{4}{}_*}\,\afp^{-1}
  \mspace{-12mu}\prod_{n=1,2,3,4}\mspace{-10mu}
  \bigl(\cosh\PSI-\sinh\PSI\cos(\PHI\!-\!\PHI_n)\bigr)^{1/2}\!\period
\end{split}\raisetag{9pt}
\end{equation}
There are no outgoing or ingoing directions along which
radiation vanishes in this case---see, e.g., Fig.~\ref{fig:dpr}(Dd).

To summarize, when $\scri$ is timelike the radiation fields
depend on direction along which the infinity is approached.
Analogously to the ${\Lambda>0}$ case \cite{KrtousPodolskyBicak:2003}
the radiation pattern has a universal
character determined by the \emph{algebraic type} of the fields.
However, new features occur when ${\Lambda<0}$:
both \emph{outgoing} and \emph{ingoing} patterns have to be studied,
their shapes depend also on the \emph{orientation of PNDs}
with respect to the infinity,
and an interesting possibility of PNDs \emph{tangent to $\scri$} appears.
Radiation vanishes only along directions which are reflections of
PNDs with respect to $\scri$,
in a \emph{generic} direction it is \emph{nonvanishing}.
The absence of ${\afp^{-1}}$ term thus cannot be used to
distinguish  nonradiative sources:
near an anti-de~Sitter-like infinity the radiative component
reflects not only properties of the sources but also
their  relation to the observer.

\begin{acknowledgments}
This work has been supported by the grants GA\v{C}R 202/02/0735 and GAUK 166/2003.
\end{acknowledgments}


\longversion{\vspace*{-15pt}}

\end{document}